\documentclass[preprint,3p,times]{elsarticle}

\usepackage{amssymb}
\usepackage{graphicx}
\journal{Physics Letters B}

\begin{document}

\begin{frontmatter}



%\begin{flushright}
%IZTECH-P-08-04
%\end{flushright}
\title{\Large Tevatron Higgs Mass Bounds: Projecting U(1)$^{\prime}$ Models to LHC Domain}

\author{Hale Sert$^{a}$, Elif Cincio\~glu$^{b}$, Durmu{\c s} A. Demir$^{a}$  and Levent Solmaz$^{b}$}

\address{$^a$ Department of Physics, {\.I}zmir Institute of
Technology, TR35430, {\.I}zmir, Turkey }

\address{$^b$ Department of Physics, Bal{\i}kesir University, TR10145, Bal{\i}kesir, Turkey}

\date{\today}

\begin{abstract}
We study Higgs boson masses in supersymmetric models
with an extra U(1) symmetry to be called U(1)$^{\prime}$.
Such extra gauge symmetries are urged by the $\mu$ problem of the MSSM,
and they also arise frequently in low-energy supersymmetric
models stemming from GUTs and strings.

We analyze mass of the lightest Higgs boson and various other
particle masses and couplings by taking into account the LEP bounds
as well as the recent bounds from Tevatron experiments. We find that
the $\mu$-problem motivated generic low-energy U(1)$^{\prime}$ model
yields Higgs masses as large as $\sim 200\ {\rm GeV}$ and violate
the Tevatron bounds for certain ranges of parameters. We analyze
correlations among various model parameters, and determine excluded
regions by both scanning the parameter space and by examining
certain likely parameter values. We also make educated projections
for LHC measurements in light of the Tevatron restrictions on the
parameter space.

We further analyze certain benchmark models stemming from E(6)
breaking, and find that they elevate Higgs boson mass into
Tevatron's forbidden band when U(1)$^{\prime}$ gauge coupling takes
larger values than the one corresponding to one-step GUT breaking.

\end{abstract}

\begin{keyword}
% \pacs{12.60.-i,14.80 Da, 14.80.Ec}
Supersymmetric U(1)$^\prime$ models, Neutral Higgs bosons, Tevatron
Higgs measurements.
\end{keyword}
\end{frontmatter}

% \linenumbers

%\section{}
\label{}


\section{Introduction}

Minimal supersymmetric model (MSSM) is the most economic extension that
can solve the naturalness problem associated with the Higgs sector of
the Standard Model of strong and electroweak interactions (SM)
\cite{Langacker:2009my}. It is an economical description since it is based
on the particle spectrum and gauge structure of the SM.
Whether it is supersymmetric or not, if the gauge structure
is extended to include new factors or embedded in a larger group
then there necessarily arise novel particle spectra and phenomena that can be
tested via collider experiments or astrophysical observations.

The simplest gauge extension of the MSSM would be to expand its
gauge group by an additional Abelian factor -- to be hereon called
$U(1)^{\prime}$ invariance. The most direct motivation for such an
extra group factor is the need to solve the $\mu$ problem of the
MSSM \cite{Kim:1983dt}. Indeed, the mass term of the Higgsinos
\begin{eqnarray}
\label{superpot-MSSM} \widehat{W}_{MSSM} \ni \mu \widehat{H}_u\cdot
\widehat{H}_d
\end{eqnarray}
involves a dimensionful parameter $\mu$ which is completely
unrelated to the soft supersymmetry breaking sector containing the
mass parameters in the theory. For consistent electroweak breaking,
the soft supersymmetry breaking mass parameters must lie at the
electroweak scale, and there is no clue whatsoever why $\mu$ should
be fixed to this very scale. For naturalizing the $\mu$ parameter a viable
approach is to associate $\mu$ to the vacuum expectation value (VEV)
of a new scalar \cite{cvetic}
\begin{eqnarray}
\label{mu-def} \mu \propto \langle S \rangle
\end{eqnarray}
where the chiral superfield $\widehat{S}$ replaces the bare $\mu$
parameter in (\ref{superpot-MSSM}) via
\begin{eqnarray}
\label{superpot} \widehat{W} \ni h_s \widehat{S} \widehat{H}_u\cdot
\widehat{H}_d
\end{eqnarray}
with $h_s$ being a Yukawa coupling. For the new superpotential not
to contain a bare $\mu$ term like (\ref{superpot-MSSM}) it is
obligatory that the U(1)$^{\prime}$ charges of the all superfields
sum up to zero by gauge invariance
\begin{eqnarray}
\label{u(1)'gauge-inv}
Q_{S} + Q_{H_u} + Q_{H_d} = 0\,.
\end{eqnarray}
Clearly, $Q_{S} \neq 0$. These conditions guarantee that a bare
$\mu$ term as in (\ref{superpot-MSSM}) is forbidden completely, and
$\mu$ parameter is deemed to arise from the VEV of $S$ via
(\ref{mu-def}).

Every single term in the superpotential satisfies U(1)$^{\prime}$
gauge invariance conditions like (\ref{u(1)'gauge-inv}). Nevertheless,
there are additional non-trivial constraints necessary to make such
models anomaly free, especially when the concerning U(1)$^\prime$
model deviates from the authentic E(6) structures. The anomalies
can be cancelled either by introducing family non-universal
charges \cite{little-hier2} or by importing novel matter species (mimicking those
of GUTs such as E(6)) (see the second reference in \cite{cvetic}). In the
present work we shall assume that anomalies are cancelled by
additional matter falling outside the reach of LHC experiments.

The $\mu$ problem detailed above is not the only motivation for introducing an
extra U(1). Indeed, such extra gauge factors, typically more than a
single U(1), arise in effective theories stemming from
supersymmetric GUTs and strings \cite{gut-string}. In such models,
the U(1) charges of fields are fixed by the unified theory. These
models are phenomenologically rich and  theoretically ubiquitous in
superstring theories and GUTs descending from SO(10) and E(6) groups
\cite{KingDiener}. The E(6) breaking pattern
\begin{eqnarray}
\label{e6-break}
E(6) \rightarrow SO(10)\otimes U(1)_{\psi} \rightarrow SU(5)\otimes
U(1)_{\chi} \otimes U(1)_{\psi} \rightarrow G_{SM}\otimes
U(1)^{\prime}
\end{eqnarray}
gives rise to the $G_{SM}\otimes U(1)^{\prime}$ model at low
energies. Each arrow in this chain corresponds to spontaneous
symmetry breakdown at a specific energy scale. Here, by
construction,
\begin{eqnarray}
\label{U1ptheta} U(1)^{\prime} = \cos\theta_{E(6)}\, U(1)_{\psi} -
\sin\theta_{E(6)}\, U(1)_{\chi}
\end{eqnarray}
is a light $U(1)^{\prime}$ invariance broken near the ${\rm TeV}$
scale whereas the other orthogonal combination $U(1)^{\prime \prime}
=\cos\theta_{E(6)}\, U(1)_{\chi} + \sin\theta_{E(6)}\, U(1)_{\psi}$
is broken at a much higher scale not accessible to LHC experiments.
The angle $\theta_{E(6)}$ designates the breaking direction in
$U(1)_{\chi} \otimes U(1)_{\psi}$ space and it is a function of the
associated gauge couplings and VEVs that realize the symmetry
breaking. Many other models can be constructed from the combination
of $\psi$ and $\chi$ models leading to a solution for $\mu$ problem
(an exception is the $\chi$ model ($\theta_{E(6)} =-\frac{\pi}{2}$)
where the singlet $S$ acquires vanishing U(1)$^{\prime}$ charge)
\cite{gut-string}.

The extra U(1) gives rise to a number of phenomena not found in the
MSSM: Its gauge boson $Z^{\prime}$ and gauge fermion
$\widetilde{Z}^{\prime}$ cause anomalies in various MSSM-specific
processes \cite{ali,zerwas}. Another point as important as these
phenomena concerns the Higgs sector: The Higgs sector of such models
differ from those of the SM and MSSM \cite{higgs-sector-SM&MSSM} not
only by the presence of extra Higgs states but also by the
modifications in the masses and couplings of the Higgs bosons
\cite{Demir:2003ke,Barger:2006dh,Hayreter:2008vr} (for
phenomenological consequences of  an extra singlet on the masses,
couplings and decay widths of Higgs bosons the reader can refer to
\cite{Barger:2006dh}). In fact, the dependencies of the Higgs masses
on the model parameters are different than in the MSSM, and the
little hierarchy problem of the MSSM seems to be largely softened in
such models \cite{little-hier1, little-hier2}.

At the wake of LHC experiments, it is convenient to study the Higgs
boson masses in U(1)$^{\prime}$ models. Apart from various mass and
coupling ranges favored by the models, the existing bounds from the
LEP and Tevatron experiments can guide one to more likely regions of
the parameter space. The LEP experiments \cite{lep} have ended with
a clear preference for the lightest Higgs boson mass:
\begin{eqnarray}
m_h > 114.4\ {\rm GeV}\,.
\end{eqnarray}
The knowledge of the Higgs mass has recently been further supported
by the Tevatron results \cite{tevatron} which state that the
lightest Higgs boson cannot have a mass in the range
\begin{eqnarray}
159\ {\rm GeV} < m_h < 168\ {\rm GeV}\,.
\end{eqnarray}
It is clear that LEP bound influences the parameter spaces of the
SM, MSSM and its extensions like NMSSM and U(1)$^{\prime}$ models.
The reason is that the LEP range is covered by all these models of
electroweak breaking. However, it is obvious that the Tevatron bound
has almost no impact on the MSSM parameter space within which $m_h$
cannot exceed $\sim 135\ {\rm GeV}$. For the same token, however,
the Tevatron bounds can be quite effective for extensions of the
MSSM whose lightest Higgs bosons can weigh above $2 M_W$. This is
the case in NMSSM not explored here and in U(1)$^{\prime}$ models
\cite{Demir:2003ke}.

In this work we shall analyze U(1)$^{\prime}$ models in regard to
their Higgs mass predictions and constrained parameter space
under the LEP as well as Tevatron bounds by assuming that the Higgs
boson searched by D$\varnothing$ and CDF corresponds to that of
the U(1)$^{\prime}$ models. In course of the analysis, we shall
consider the U(1)$^{\prime}$ model achieved by low-energy
considerations as well as by high-energy considerations (the
GUT and stringy U(1)$^{\prime}$ models mentioned above). In each
case we shall scan the parameter space to determine the bounds
on the model parameters by imposing the bounds from direct
searches.

The rest of the paper is organized as follows: In Sec. II below we
discuss certain salient features of the U(1)$^{\prime}$ models in
regard to collider bounds on $M_{Z^{\prime}}$. Sec. III is devoted
to a detailed analysis of the U(1)$^{\prime}$ models selected. In
Sec. IV we conclude.

\section{Phenomenological Aspects of U(1)$^\prime$ Models}
In this section we provide a brief overview of the fundamental
constraints on U(1)$^{\prime}$ model. First of all, the
U(1)$^{\prime}$ model is known to generate the neutrino masses in
the correct experimental range via Dirac type coupling. The scalar
field $S$ responsible for generating the $\mu$ parameter also
generates the neutrino Dirac masses \cite{lisa1}. Furthermore, the
same model offers a viable cold dark matter candidate via the
lightest right-handed sneutrino, and accounts for the PAMELA and
Fermi LAT results for positron excess for a reasonable set of
parameters \cite{ismail1}. Hence, there is no reason for insisting
that the neutralino sector offers a CDM candidate. Our focus in this
work is on the Higgs sector to which neutrino sector gives no
significant contribution.

 An important point which concerns the anomalies. A generic U(1)$^{\prime}$
model suffers from triangular anomalies and hence gauge coupling
non-unification. In the E(6)-motivated models, by construction, all
anomalies automatically cancel  out when the complete E(6)
multiplets are included. For a generic U(1)$^{\prime}$, with minimal
matter spectrum, cancellation is nontrivial. One possibility is to
introduce U(1)$^{\prime}$ models with family-dependent charges
\cite{little-hier2}. Another possibility is that anomalies are
cancelled by heavy states (beyond the reach of LHC) weighing near the {\rm TeV} scale.
We shall follow this possibility.

 The Higgs sector of the model, as mentioned before, involves the
singlet Higgs $S$ and the electroweak doublets $H_u$ and $H_d$. All
of them are charged under U(1)$^{\prime}$ gauge group. The Higgs
fields expand around the vacuum state as follows
\begin{eqnarray}
\label{higgsexp} H_u  = \frac{1}{\sqrt{2}}\left(\begin{array}{c}
\sqrt{2} H_u^{+}\\ v_u + \phi_u + i \varphi_u\end{array}\right)
\:,\:\:~~ H_d  = \frac{1}{\sqrt{2}}\left(\begin{array}{c} v_d +
\phi_d + i \varphi_d\\ \sqrt{2} H_d^{-}\end{array}\right)\:,\:\:
 ~~S = \frac{1}{\sqrt{2}} \left(v_s + \phi_s +
i \varphi_s\right),
\end{eqnarray}
where $H_{u}^{+}$ and $H_d^{-}$ span the charged sector involving
the charged Goldstone eaten up by the $W^{\pm}$ boson as well as the
charged Higgs boson. The remaining ones span the neutral degrees of
freedom: $\phi_{u,d,s}$ are scalars and $\varphi_{u,d,s}$ are
pseudoscalars. In the vacuum state
\begin{eqnarray}
\frac{v_{u}}{\sqrt{2}} \equiv \langle H_u^0 \rangle\,,\;~~
\frac{v_d}{\sqrt{2}} \equiv \langle H_d^0\rangle\,,\;~~
\frac{v_s}{\sqrt{2}} \equiv \langle S \rangle
\end{eqnarray}
the $W^{\pm}$, $Z$ and $Z^{\prime}$ bosons all acquire masses.
However, the neutral gauge bosons $Z$ and $Z^{\prime}$ exhibit
nontrivial mixing \cite{Langacker:2008yv,cvetic}. The two
eigenvalues of this mixing matrix \cite{Langacker:2008yv} give the
masses of the physical massive vector bosons ($M_{Z_1}, M_{Z_2}$)
where $M_{Z_1}$ must agree with the experimental bounds on the $Z$
boson mass in the MSSM (or SM). The mixing angle $\alpha_{Z-Z'}$
\cite{Langacker:2008yv} must be a few $10^{-3}$ for precision
measurements at LEP experiments to be respected. This puts a bound
on the $Z_2$ boson mass. In particular, in generic E(6) models
$m_{Z_2}$ must weigh nearly a ${\rm TeV}$ or more according to the
Tevatron measurements \cite{erler,tevatron2}.

 Due to the soft breaking of supersymmetry, the Higgs boson masses
shift in proportion to particle--sparticle mass splitting under
quantum corrections. Though all particles which couple to the Higgs
fields $S$, $H_{u}$ and $H_{d}$ contribute to the Higgs boson
masses, the largest correction comes from the top quark and its
superpartner scalar top quark (and to a lesser extent from the
bottom quark multiplet). Including  top and bottom quark
superfields, the superpotential takes the form
\begin{eqnarray}
\widehat{W} \ni h_s \widehat{S} \widehat{H}_{u}\cdot \widehat{H}_d +
h_t \widehat{Q}\cdot \widehat{H}_u \widehat{t}_R^c+ h_b
\widehat{Q}\cdot \widehat{H}_d \widehat{b}_R^c
\end{eqnarray}
where $h_t$ and $h_b$ are top and bottom Yukawa couplings. Clearly
$\widehat{Q}^T = \left( \widehat{t}_L, \widehat{b}_L\right)$. This
superpotential encodes the dominant couplings of the Higgs fields
which determine the $F$--term contributions.

Effective potential proves to be an efficient method for computing
the radiative corrections to Higgs potential. In fact, the
radiatively corrected potential reads as
\begin{eqnarray}
\label{total-pot} V_{total}\left(H\right) = V_{tree}\left(H\right) +
\Delta V\left(H\right)
\end{eqnarray}
where the tree level potential is composed of $F$--term, $D$--term
and soft--breaking pieces
\begin{equation}
V_{tree} = V_F + V_D + V_{soft}
\end{equation}
with
\begin{eqnarray}
&&V_F  = |h_s|^2 \left[|H_u \cdot H_d|^2 + |S|^2 (|H_u|^2 + |H_d|^2)\right]\,,\\
&&V_D  = \frac{G^2}{8}  \left(|H_u|^2 - |H_d|^2 \right) +
\frac{g_2^2}{2} \left( |H_u|^2 |H_d|^2 - |H_u \cdot H_d|^2 \right )
+ \frac{{g^\prime_Y}^2}{2} \left (Q_{H_u}
|H_u|^2 + Q_{H_d} |H_d|^2 + Q_S |S|^2 \right )^2\,,\\
&&V_{soft} = m_{H_u}^2 |H_u|^2 + m_{H_d}^2 |H_d|^2 + m_s^2|S|^2 + (
h_s A_s S H_u \cdot H_d + h.c.)\,.
\end{eqnarray}

The contributions of the quantum fluctuations in (\ref{total-pot})
read as
\begin{eqnarray}
\label{deltaV} \Delta V = \frac{1}{64\,\pi^2}
\mbox{Str}\left[\mathcal{M}^4\left(\ln
\frac{\mathcal{M}^2}{\Lambda^2}-\frac{3}{2}\right)\right]
\end{eqnarray}
where $\mbox{Str}\equiv\sum_J (-1)^{2 J}(2J+1) \mbox{Tr}$ is the
usual supertrace which generates a factor of $6$ for squarks and
$-12$ for quarks. $\Lambda$ is the renormalization scale  and
$\mathcal{M}$ is the field-dependent mass matrix of quarks and
squarks (we take $\Lambda= m_t + m_{Z_2}/2$). The dominant
contribution comes from top quark (and bottom quark, to a lesser
extent) multiplet. The requisite top and bottom quark
field-dependent masses read as $m_t^2\left(H\right) = h_t^2
\left|H_u^0\right|^2$,  $m_b^2\left(H\right) = h_b^2
\left|H_d^0\right|^2$. The mass-squareds of their superpartners
follow from
\begin{eqnarray}
\label{stopmass} m^2_{\tilde{f}} = \left(\begin{array}{c c}
 M^2_{\tilde{f}LL}& M^2_{\tilde{f}LR}\\
 M^{2}_{\tilde{f}RL}& M^2_{\tilde{f}RR}
\end{array}\right)
\end{eqnarray}
where $f=t$ or $b$. For instance, the entries of the stop
mass-squared matrix read to be
\begin{eqnarray}
\label{entries}
M^2_{\tilde{t}LL} &=& m^2_{\tilde{Q}} + m^2_t- \frac{1}{12}
\left(3g_2^2-g_Y^2\right) (|H_u^0|^2 - |H_d^0|^2)+g^{\prime2}_Y\,
Q_Q\left(Q_{H_u} |H_u|^2+Q_{H_d} |H_d|^2+Q_S |S|^2\right)
\nonumber\\
M^2_{\tilde{t}RR} &=& m^2_{\tilde{t}_R} + m^2_t-\frac{1}{3}g_Y^2
(|H_u^0|^2 -|H_d^0|^2)+g^{\prime2}_Y\, Q_U\left(Q_{H_u}
|H_u|^2+Q_{H_d} |H_d|^2+Q_S |S|^2\right)
\nonumber\\
 M^2_{\tilde{t}LR} &=& M^2_{\tilde{t}RL} = h_t\left( A_t {H^{0}_u}-h_s S {H^0_d}\right)
\end{eqnarray}

Insertion of the top and bottom mass matrices into (\ref{deltaV})
generates the full one-loop effective potential. Radiatively
corrected Higgs masses and mixings are computed from the effective
potential \cite{Demir:2003ke}.

\section{Analysis}
In this section we shall perform a numerical analysis of Higgs boson
masses in order to determine the allowed regions under the LEP and
Tevatron bounds. Our results, with a sufficiently wide range for
each parameter, can shed light on the relevant regions of the
parameter space to be explored by the experiments at CERN. In the
following we will first discuss the parameter space to be employed,
and then we shall provide a set of figures each probing certain
parameter ranges in the U(1)$^{\prime}$ models considered.

\subsection{Parameters}
In course of the analysis, we shall partly scan the parameter space and
partly analyze certain parameter regions which best exhibit the bounds from
the Higgs mass measurements. We first list down various parameter values
to be used in the scan.

 {\it U(1)$^{\prime}$ Gauge Coupling.} The U(1)$^{\prime}$ models we consider are inherently
{\it unconstrained} in that, irrespective of their low--energy or high-energy origin, we let
U(1)$^{\prime}$ gauge coupling ${g^{\prime}_Y}$ to vary in a reasonable range in units of
the hypercharge gauge coupling. We thus call all the models we investigate as `Unconstrained
U(1)$^{\prime}$ Models', or, UU(1)$^{\prime}$ Models, in short.

We shall be dealing with four different UU(1)$^\prime$ models:
\begin{itemize}
\item UU(1)$^{\prime}$ from E(6) supersymmetric GUT:
The $\eta$, $N$ and $\psi$ Models.

\item UU(1)$^{\prime}$ from low-energy (solution of the $\mu$ problem):
This is the low-energy model obtained by taking
$Q_{H_u}=Q_{H_d}=Q_{Q}=-1$ and hence $Q_U=Q_D=Q_S=2$, and we shall
be calling this model the $X$ Model.
\end{itemize}
The charge assignments of E(6)-based models can be found in
\cite{Langacker:2008yv}. For them we use the same symbols but mutate
them by giving up the typically-assumed value ${g^{\prime}_Y} =
\sqrt{\frac{5}{3} (g_2^2 + g_Y^2)} \sin\theta_W$ (obtained by
one-step GUT breaking), and changing it in the range $g_Y$ to $2
g_Y$. The motivation behind this mutation of the E(6)-based
U(1)$^{\prime}$ groups is that one-step GUT breaking is too
unrealistic to follow; the GUT group is broken at various steps as
indicated in (\ref{e6-break}). Nevertheless, large values of
$g_Y^{\prime}$ may be inadmissibly large for perturbative dynamics,
and we shall note this feature while interpreting the figures.
Despite this, however, by varying the $g_Y^{\prime}$ we will treat
E(6)-based models as some kind of specific UU$^\prime$ models in
which we can probe the impact of different $g^{\prime}_Y$ values on
the lightest Higgs mass.

Unlike the E(6)-based models, we adopt the value of $g_Y^{\prime}$
from one-step GUT breaking in analyzing the $X$ model. In $X$
model, by the need to cancel the anomalies, we assume that
there exist an unspecified sector of fairly light chiral
fields, and normalization of the charge and other issues
depend on that sector \cite{cvetic}. Our analysis will
be indicative of a generic U(1)$^{\prime}$ model stemming
from mainly the need to evade the naturalness problems
associated with the $\mu$ problem of the MSSM.

{\it The Gauge and Yukawa Couplings.} In U(1)$^\prime$ models, at
the tree level
  one can write $m^2_h\lesssim\,a_i+b_i\,h^2_s$
  where $a_i,b_i$ are some constants to be determined from
  the given value of $\tan\beta$, charge assignments as well as the
  soft supersymmetry-breaking sector. Hence, for sufficiently large
  $b_i/a_i$ ratios, one can expect $m_h\propto\,h_s$. At one-loop level, it is interesting to
probe if such a relation also exists for the gauge coupling,
Yukawa coupling and other important model parameters. We will
be dealing with this issue numerically, by changing the value of
$g^{\prime}_Y$ as stated above.

{\it The Z-Z$^{\prime}$ Mixing.} We shall always require the
$Z-Z^{\prime}$ mixing to obey the bound
$|\alpha_{Z-Z^\prime}|<10^{-3}$ for consistency with current
measurements \cite{Abazov:2008xq}. The collider analyses
\cite{tevatron2} constrain $m_{Z_2}$ to be nearly a ${\rm TeV}$ or
higher with the assumption that $Z_2$ boson decays exclusively into
the SM fermions. However, inclusion of decay channels into
superpartners increases the $Z_2$ width, and hence, decreases the
$m_{Z_2}$ lower bound by a couple of $100\ {\rm GeV}$s
\cite{Langacker:2008yv}. But, for simplicity and definiteness, we
take $ m_{Z_2} \geq 1\ {\rm TeV}$ as a nominal value.

 {\it Ratio of the Higgs VEVs $\tan\beta$.} We fix $\tan\beta$ from the knowledge of $\alpha_{Z-Z^\prime}$ \cite{Demir:2003ke}: ~~~ $ \tan^2\beta = F_d/F_u$ ~~~~
  where
\begin{equation}
F_{u,(d)} = (2 g^{\prime}_Y/G) Q_{H_{u,(d)}} \pm \alpha_{Z-Z^\prime}
(-1 + (2 g^{\prime}_Y/G)^2 ( Q_{H_{u,(d)}}^2 + Q_S^2
(v_s^2/v^2)))\,.
\end{equation}
Using this expression we find that $\tan\beta$ stays around 1 (this
is true as far as $v_s$ is not very large), and thus, we scan
$\tan\beta$ values from $0.5$ to $5$ in E(6)-based models, and  in
the $X$ Model. The post-LEP analyses of the MSSM disfavors
$\tan\beta\sim 1$ yet in U(1)$^{\prime}$ models as well as in NMSSM
there is no such conclusive result. One can in fact, consider
$\tan\beta$ values significantly smaller than unity, as a concrete
example $\eta$ model favors $\tan\beta=0.5$.

 {\it The Higgsino Yukawa Coupling.} Our analysis respects $h_s = 1/\sqrt{2}$ in our $X$ model; this value is suggested by the RGE analysis of
\cite{cvetic}. However, not only for our $X$ model but also for our
mutated E(6) models  we allow $h_s$ to vary from $0.1$ to  $0.8$ for
determining its impact on the Higgs boson masses. The Higgsino Yukawa
coupling $h_s$ determines the effective $\mu$ parameter in units of
the singlet VEV $v_s$.

 {\it The Squark Soft Mass-Squareds.} We scan each of $m_{\tilde{Q}}, m_{\tilde{t}_R}$ and   $m_{\tilde{b}_R}$ in $[0.1,1]\ {\rm TeV}$ range.
Following the PDG values \cite{PDG}, we require light stop and sbottom to weigh appropriately: $m_{\tilde{t}_1}>180\ {\rm GeV}$ and $m_{\tilde{b}_1}>240\ {\rm GeV}$. These bounds follow from direct searches at the Tevatron and other colliders.

 {\it Singlet VEV $v_s$.} We scan $v_s$ in $[1,2]\ {\rm TeV}$ range so that $m_{Z_2}$ can
be larger than $1\ {\rm TeV}$. In doing this we set
$\mu_{eff}<1\,{\rm TeV}$ as the upper limit of this parameter.
Larger values of $\mu_{eff}$ are more fine-tuned in such models than
the MSSM \cite{Barger:2006dh}. Such keen values of $v_s$ and
$\mu_{eff}$ turn out to be necessary for keeping the mentioned
models at the low energy region and also for satisfying the
aforementioned constraints.

 {\it Trilinear Couplings.} In the general scan we vary each of $A_t$, $A_b$, $A_s$ in $[-1,1]\ {\rm TeV}$ range,
independently. This is followed by a specific scan regarding
Tevatron bounds where the trilinears and soft masses of the scalar
quarks are assigned to share some  common values.
 We do this for all of the models we are considering.

These parameter regions will be employed in scanning the parameter space for
determining the allowed domains. In addition to and agreement with these,
we shall select out certain parameter values to illustrate how strong or
weak the bounds from Higgs mass measurements can be. The results are displayed
in a set of figures in the following subsection.

\subsection{Scan of the Parameter Space}

In this subsection we present our scan results for various model
parameters in light of the Tevatron and LEP bounds on the lightest
Higgs mass. We start the analysis with a general scan using the
inputs mentioned in the previous subsection. This will allow us to
perform  a specific search  concentrated around the Tevatron
exclusion limits. In both of the scans we will present the results
for  $X$ model first, which is followed by the E(6)-based models
$\eta$, $N$ and $\psi$ models.

\begin{figure}[ht]
\begin{center}
\includegraphics[scale=1,height=11.5cm,angle=0]{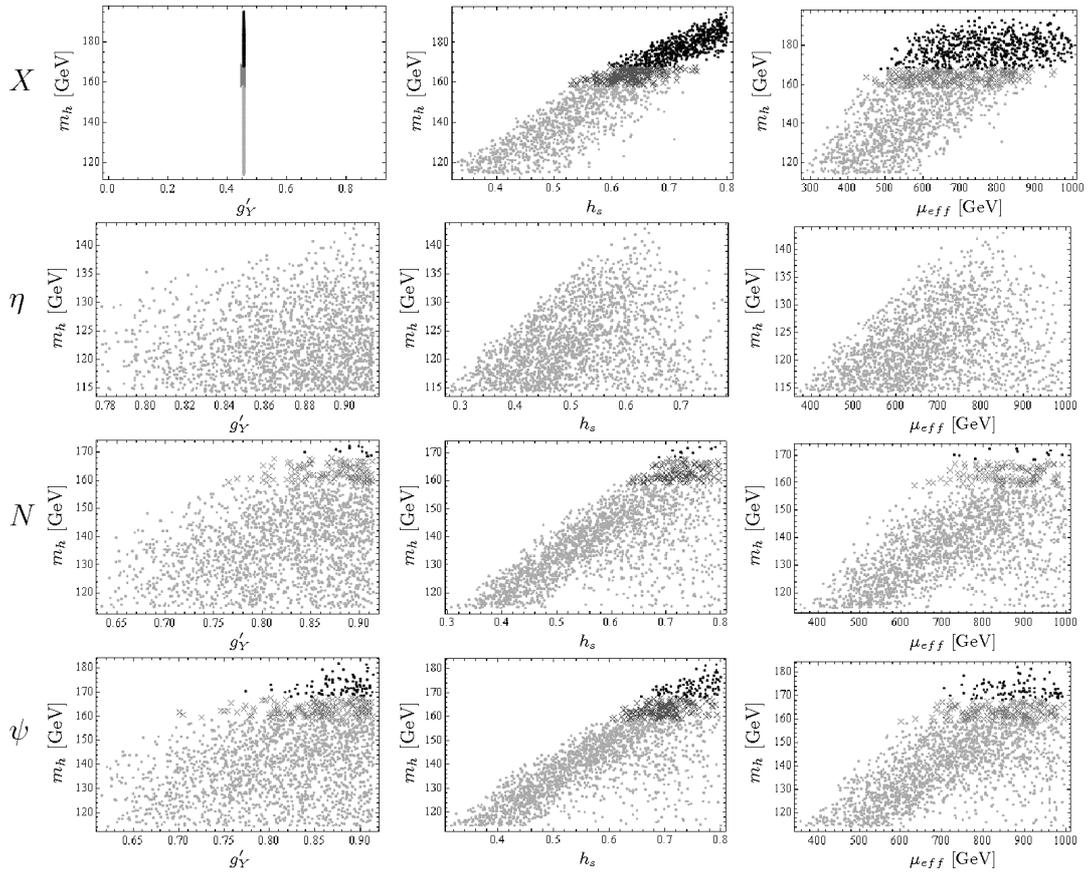}
\end{center}
\caption{The plots for the $X$,$\eta,N$ and $\psi$ models (from top
to bottom). The mass of the lightest Higgs boson against the gauge
coupling $g^\prime_Y$ (left-panels), Higgsino Yukawa coupling $h_s$
(middle-panels), and effective $\mu$ parameter (right-panels). The
shading convention is such that the points giving $m_h
> 168\ {\rm GeV}$ are shown by black dots,  those yielding $114.4\ {\rm
GeV}\leq m_h \leq 159\ {\rm GeV}$ by grey dots,  and those yielding
$159\ {\rm GeV}\leq m_h \leq 168\ {\rm GeV}$ by  grey crosses.}
\label{fig1}
\end{figure}

Related with the general scan we present  Fig. \ref{fig1} wherein
$h_s$, $g^\prime_Y$ and $\mu_{eff}$ are variables on the surface
(The only exception is $X$ model for which $g^\prime_Y$ is taken at its GUT normalized value.).
The remaining variables, whose ranges were mentioned in the previous section, vary
in the background.  In  Fig. \ref{fig1}, shown are the variations of
the lightest Higgs boson mass against the gauge coupling
$g^\prime_Y$ (left-panels), Higgsino Yukawa coupling $h_s$
(middle-panels), and the effective $\mu$ parameter $\mu_{eff}$
(right-panels).

As are seen from the left panels of Fig. \ref{fig1}, increase in the
$g^\prime_Y$ gives rise to higher upper bounds on $m_h$ for E(6)-based models.
The same behavior, though not shown explicitly, occurs in the $X$ model
(which already yields $m_h$ values as high as $195\ {\rm GeV}$). Excepting
the $\eta$ model, the E(6)-based models are seen to accommodate Higgs
boson masses larger than the Tevatron upper bound when $g_Y^{\prime}$
rises to extreme values above $\sim 0.8$. Needless to say, the regions
with grey dots are followed by regions with grey crosses (the forbidden region),
as expected from the dependence of the Higgs boson mass on $g_Y^{\prime}$.
The $\eta$ model does not touch even the Tevatron lower bound of the excluded
region for the parameter values considered.

Depicted in the middle panels of Fig. \ref{fig1} is the variation
of the Higgs boson mass with the Higgsino Yukawa coupling for the
models considered. Clearly, $h_s$ parameter is more determinative
than $g_Y^{\prime}$ in that $m_h$ tends to stay in a strip of
values for the entire range of $h_s$. Indeed, upper bound on $m_h$ (and
its lower bound, to a lesser extent) varies linearly with $h_s$ for $X,N$ and $\psi$ models.
This is also true for the $\eta$ model at least up to $h_s\sim 0.65$. In general,
Tevatron bounds divide $h_s$ values into two disjoint regions
separated by the forbidden region yielding $m_h$ values excluded
by the Tevatron results. One keeps in mind that, in this and following figures,
the $\eta$ model serves to illustrate E(6)-based models yielding
a genuine light Higgs boson: The Higgs boson stays light for the entire
range of parameter values considered. At least for the $X$ model, one
can write
\begin{equation}
159\gtrsim\,m_h \gtrsim\,114.4\,\,\Rightarrow\,h_s\,\,\in
[0.3,0.7]\,\,\, \mbox{and}\,\,\,m_h \gtrsim\,168\,\,\Rightarrow\,
h_s\,\,\in [0.6,0.8]
\end{equation}
from the distribution of the allowed regions (top middle panel).
More precisely, the Higgsino Yukawa coupling largely determines the
ranges of the Higgs mass in that while $m_h$ barely saturates the
lower edge of the Tevatron exclusion band for $h_s<0.52$, it takes
values above the Tevatron upper edge for $h_s > 0.58$. In other
words, Tevatron bound divides $h_s$ ranges into two regions in
relation with $m_h$ values: The $h_s$ values for low $m_h$ ( $114.4\
{\rm GeV}\leq m_h \leq 158\ {\rm GeV}$) and those for high $m_h$
($m_h > 168\ {\rm GeV}$). This distinction is valid for all the
variables we are analyzing.

The variation of the Higgs boson mass with the effective $\mu$ parameter
is shown in the right-panels for Fig. \ref{fig1}, for each model.
It is clear that $\mu_{eff}\gtrsim\,300\,{\rm GeV}$  for the LEP
bound to be respected. On the other hand, one needs
$\mu_{eff}\gtrsim\,500\,{\rm GeV}$  for $m_h$ to touch the lower
limit of the Tevatron exclusion band in the $X$ model. Similar conclusions
hold also for the mutated E(6) models: $\mu_{eff}\gtrsim\ 700\ {\rm GeV}$
for $\psi$ and $N$ models (while the forbidden Tevatron territory is never
reached in the $\eta$ model). The $\eta$ model is bounded by LEP data only
(at least within the input values assumed for which we considered
 $v_s\leq\,2$ TeV).

From the scans above we conclude that:
\begin{itemize}
\item All models are constrained by the LEP bound, that is, each
of them predict Higss masses below $114.4\ {\rm GeV}$ for certain
ranges of parameters.

\item The $X$ model, a genuine low-energy realization of UU(1)$^{\prime}$
models based solely on the solution of the $\mu$ problem, yields
large $m_h$ values, and thus, violated the Tevatron forbidden band
low values of $g_Y^{\prime}$, $h_s$ and $\mu_{eff}$ compared to the
mutated E(6)-based models. The latter require typically large values
of $g_Y^{\prime}$, $h_s$ and $\mu_{eff}$ for yielding $m_h$ values
falling within the Tevatron territory ( Meanwhile, this can happen
only if $g^\prime_Y\gtrsim\,0.77$ in $N$ model and
$g^\prime_Y\gtrsim\,0.7$ in $\psi$ model with a Yukawa coupling
saturating $h_s\gtrsim\,0.62$). In fact, the $\eta$ model does not
even approach to the $159\ {\rm GeV}$ border so that it does not
feel Tevatron bounds at all. There is left only a small parameter
space wherein $m_h$ exceeds  $159\ {\rm GeV}$ for $\psi$ and $N$
models. One can safely say that for `small' $g_Y^{\prime}$ and $h_s$
the E(6)-based models predict $m_h$ to be low, significantly below
$159\ {\rm GeV}$. In other words, Tevatron bounds shows tendency to
rule out non-perturbative behavior of E(6)-based models.

\item One notices that heavy Higgs limit typically require large $\mu_{eff}$
(close to ${\rm TeV}$ domain) and thus one expects Higgsinos to be significantly
heavy in such regions. The LSP is to be dominated by the gauginos, mainly. In
such regions, one expects the physical neutralino corresponding to $\widetilde{Z}^{\prime}$
to be also heavy due to the fact that $\widetilde{Z}^{\prime}$ mixes with $\widetilde{S}$
by a term proportional to $h_s v_s$ \cite{ali}. Therefore, the light neutralinos
are to be dominantly determined by the MSSM gauginos.
\end{itemize}

Using the grand picture reached above, we now perform a point-wise
search aiming to cover critical points wherein Tevatron exclusion
is manifest. We project implications of these exclusions to scalar fermions and other
neutral Higgs bosons. But, for doing this we first fix certain
variables, and by doing so, we get rid of overlapping regions (seen
in surface parameters while others running in the background).

From Fig. \ref{fig1}, we find it sufficient to consider values
around $h_s\sim 0.7$ and $g^\prime_Y \sim\,2 g_Y$. More precisely,
we consider Higgsino Yukawa couplings as $h_s=0.65,0.5, 0.7$ and $0.7$
for $X,\eta,N$ and $\psi$ models, respectively. We set $g^\prime_Y  = 1.9 g_Y$
for all three mutated E(6) models, while we keep it as in Fig. \ref{fig1}
for the $X$ model.

\begin{figure}[ht]
\begin{center}
\includegraphics[scale=1,height=11.5cm,angle=0]{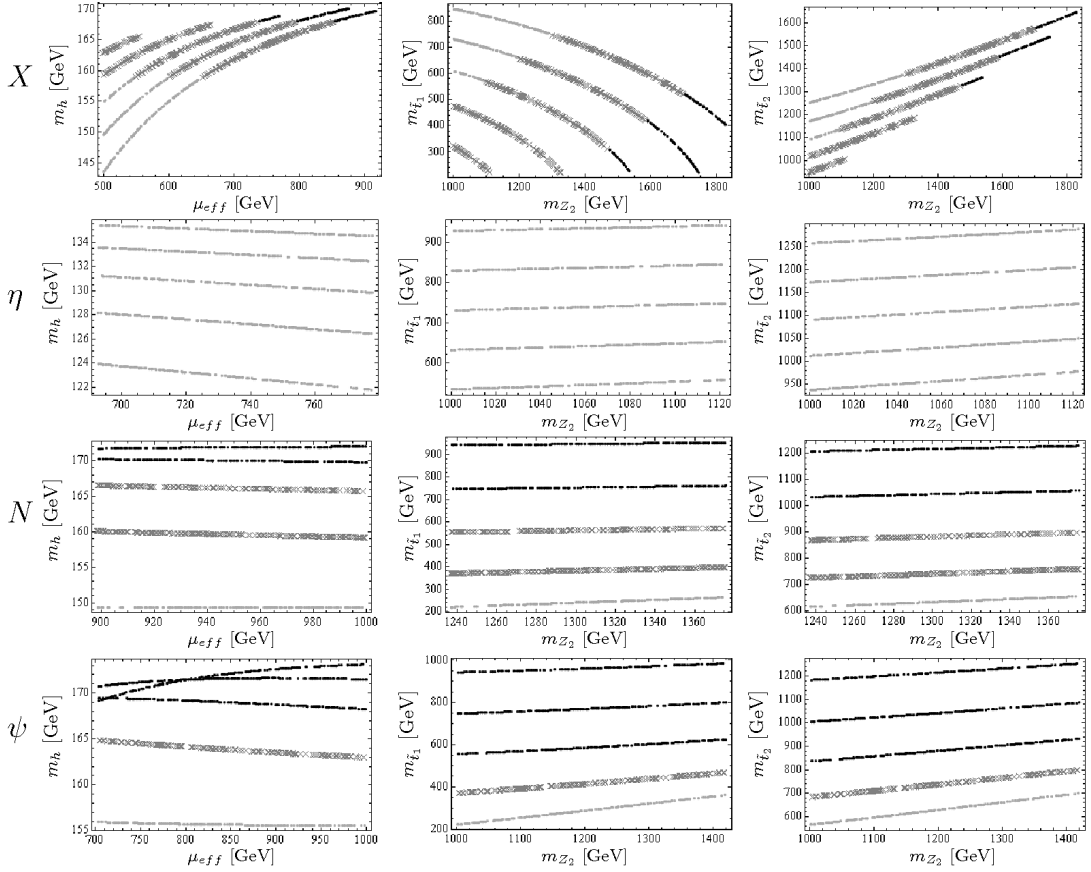}
\end{center}
\caption{The mass of the lightest Higgs boson against  the effective
$\mu$ parameter (left-panels), the mass of the light scalar top
$m_{\tilde{t}_1}$ against the mass of the ${Z_2}$ boson
(middle-panels), and the mass of the heavy scalar top
$m_{\tilde{t}_2}$ against the mass of the ${Z_2}$ boson
(right-panels) in $X,\eta,N$ and $\psi$ models (top to bottom). Our
shading convention is the same as in Fig. \ref{fig1}. The inputs
are selected as: $m_{common}=m_{\tilde{Q}}=m_{\tilde{t}_R}=m_{\tilde{b}_R}=-A_t=-A_b=-A_s=0.2$
to $1\,{\rm TeV}$ with increments 200 GeV  in $N$ and $\psi$ models. In
$X$ and $\eta$ models we scan $m_{common}$ from 0.5 to
$1\,{\rm TeV}$ with increments 100 GeV. These inputs are also used
in the following figure. In any panel of the figures we observe a
hierarchy such that largest $m_{common}$ value corresponds to the
largest $m_h$ value (topmost data lines) which is fixed at $1\ {\rm TeV}$.}
\label{fig2}
\end{figure}

In Fig. \ref{fig2}, depicted are variations of the $m_h$ and scalar
top quark masses ($m_{\tilde{t}_1}$ and $m_{\tilde{t}_2}$) with
$\mu_{eff}$ and $M_{Z_2}$. This is the targeted search. Now, as can
be seen from the left panels of Fig. \ref{fig2}, the effective $\mu$
parameter should satisfy  $\mu_{eff}>500$ GeV in $X$ model, while
others demanding higher values. This is due to already fixed $h_s$
parameter value. In this figure, the impact of Tevatron exclusions
is seen clearly (gray-crosses) on scalar fermions (middle and right
panels of $X,N$ and $\psi$ models), too. It is interesting to check
model dependent issues for this sector because the scalar fermions
shall be important for discriminating among the supersymmetric
models (even among the U(1)$^\prime$ models) at the LHC and ILC. The
goal of  Fig. \ref{fig2} is to serve this aim, in which scalar quark
masses are plotted against varying $Z_2$ boson mass (middle and
right-panels). The correlation between sfermion masses and $M_{Z_2}$
comes mainly from the U(1)$^{\prime}$ $D$-term contributions
(proportional to $g^2_{Y^{\prime}} v_s^2$) to the $LL$ and $RR$
entries of the sfermion mass-squared matrices. There are also F-term
contributions proportional to $h_s v_s$ to $LR$ entries but their
effects are much smaller compared to those in the $LL$ and $RR$
entries (see Eq. (\ref{entries}) for details). This is an important
effect not found in the minimal model: Variation of sfermion masses
with $\mu$ probes only the $LR$ entry in the MSSM. It is in such
extensions of the MSSM that one finds explicit dependence on
$\mu_{eff}$ in not only the $LR$ entries but also in $LL$ and $RR$
entries; effects of $\mu_{eff}$ are more widespread than in the
minimal model where $\mu$ is regarded as some external parameter
determined from the electroweak breaking condition.

From Fig. \ref{fig2} one concludes that variations of $m_h$ and
$m_{\tilde{t}_{1,2}}$ are much more violent in $X$ model than in the
E(6)-based models. In the $X$ model changes in $M_{Z_2}$ and
$\mu_{eff}$ influence Higgs and stop masses violently so that
allowed and forbidden regions are seen rather clearly. In E(6)-based
models what we have nearly constant strips, and thus, $m_h$ and
$m_{\tilde{t}_{1,2}}$ remain essentially unchanged with $\mu_{eff}$
and $M_{Z_2}$. Moreover, in mutated E(6) models the forbidden
regions and allowed regions fall into distinct strips, signalling
thus the aforementioned near constancy of the Higgs and stop masses.
From Fig. \ref{fig2} it is possible to read out certain likely
ranges for stop and Higgs boson masses, which will be key observables
in collider experiments like LHC and ILC. Indeed, in $X$ model one
deduces that
\begin{itemize}
\item Higgs in low-mass region\ $\Longrightarrow$\ $m_{\tilde{t}_1}\in [600,800]\ {\rm GeV}$\ and $m_{Z_2} \in [1.0,1.3]\ {\rm TeV}$,
\item Higgs in high-mass region\ $\Longrightarrow$\ $m_{\tilde{t}_1}\in [200,550]\ {\rm GeV}$\ and $m_{Z_2} \in [1.5,1.8]\ {\rm TeV}$.
\end{itemize}
Therefore, in principle, taking the $X$ model as the underlying setup,
one can determine if Higgs is in the low- or high-mass domains by a
measurement of the scalar top quark masses. For instance, if collider
searches exclude low-mass light stops up to $\sim 600\ {\rm GeV}$ then
one immediately concludes that the Higgs boson should be light, {\it i. e.}
below $ 2 M_W$.

Contrary to model $X$, E(6)-based models $N$ and $\psi$ allow the
$Z^\prime$ mass to be more confined, {\it i.e.} the mass of the
$Z_2$ boson is in $\sim[1,1.4]\ {\rm TeV}$ range within these two
models. Furthermore, these two models can rule out $m_{\tilde{t}_1}$
around $\sim\,[300,500]\ {\rm GeV}$ (One keeps in mind, however,
that in these models low (high) stop mass values are related with
low (high) $m_h$ values, in contradiction with the $X$ model).
Besides this, all three of $X$, $N$ and $\psi$ models exploration of
{\it high-mass} region demands larger values for $m_{\tilde{t}_2}$.
One notices that largest (smallest) splitting between
$m_{\tilde{t}_2}$ and $m_{\tilde{t}_1}$ is observed in $X$ ($\psi$)
model.
\begin{figure}[ht]
\begin{center}
\includegraphics[scale=1,height=11.5cm,angle=0]{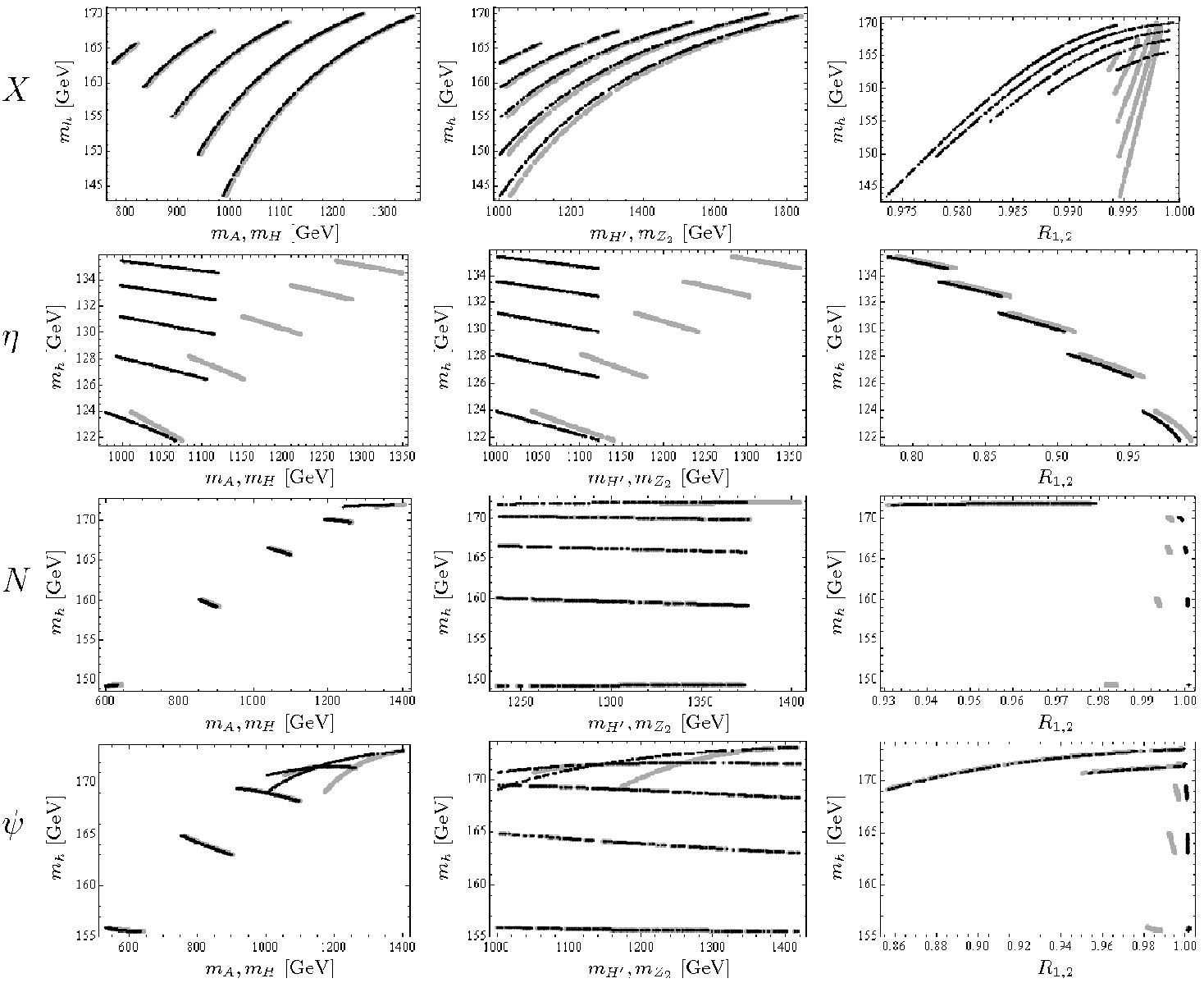}
\end{center}
\caption{Variations of the lightest Higgs boson mass $m_h$ with those
of the heavy CP-even Higgs scalars $H$, $H^{\prime}$ and of the CP-odd
scalar $A$. Also given is the dependence of $m_h$ on the $Z_2$ boson
mass. In the decoupling region, $m_H\sim m_A$ and $m_H^{\prime} \sim m_{Z_2}$.
The notation is such that $m_A$ and $m_{H^{\prime}}$ are denoted by grey dots, $m_H$ and
$m_{Z_2}$ by  black dots. As a measure of the approach to the decoupling
region, we explore, in the right-panels, the quantities $R_1$ (gray dots) and
$R_2$ (black dots). The input parameters are taken as in Fig. \ref{fig2}.} \label{fig3}
\end{figure}
As an extension of the MSSM, the present model predicts 3 CP-even
Higgs bosons: $h$, $H$ and $H^{\prime}$. There is no analogue of
$H^{\prime}$ in the MSSM. The mode predicts one single pseudoscalar
Higgs boson $A$ as in the MSSM. In the decoupling regime {\it i. e.}
when heavier Higgs bosons decouple from $h$ one expects the mass
hierarchy $m_{H^{\prime}} \sim m_{Z_2} \gg m_H \sim m_A \gg m_h$. It
is thus convenient to analyze the model in regard to its Higgs mass
spectra to determine in what regime the model is working. To this
end, we depict variations of $m_h$ with $m_H$, $m_H^{\prime}$ and
$m_{Z_2}$ in Fig. \ref{fig3}. For quantifying the analysis we define
the ratios $R_1\equiv \frac{m_H}{m_A}$, $R_2 \equiv
\frac{m_{Z_2}}{m_{H^\prime}}$ which are, respectively, shown by gray
and black dots in Fig. \ref{fig3}.

In Fig. \ref{fig3}, shown in the leftmost column are variations of $m_h$ with $m_H$ (black dots)
and with $m_A$ (grey dots). It is clear that, the $X$ and $N$ models are well inside the decoupling
regime for the parameter ranges considered. On the other hand, the $\psi$ and $\eta$ models, especially the $\eta$ model,
are far from their decoupling regime. In this regime, the lightest Higgs can weigh well above its lower bound.
One notices that, $A$ and $H$ bosons exhibit no sign of degeneracy in the $\eta$ model.

The variations of $m_h$ with $m_H^{\prime}$ and $m_{Z_2}$ are shown in the middle column of Fig. \ref{fig3}. One observes
that grand behavior is similar to those in the first column. One, however, makes the distinction that $m_h$ depends
violently on $m_H^{\prime}$ and $m_{Z_2}$ in $X$ and $\eta$ models while it stays almost completely independent for
$\psi$ and $N$ models.

All the properties summarized above are quantified in the third column wherein $m_h$ is plotted against $R_1$ and $R_2$.
The degree to which $R_{1,2}$ measure close to unity give a quantitative measure of how close the parameter values are to the
decoupling regime. One notices that they differ significantly from unity in $\eta$ and $\psi$ models.
In summary,  $m_A$/$m_H$ ratio drops to $\sim\,0.8$ in $\eta$ model. This is also true for $m_{Z_2}/m_{H^{\prime}}$.
It is interesting to observe that $R_1$ and $R_2$ behave very similar in most of the parameter space. This
figure depicts the heavy model dependency of neutral Higgs masses.

Experiments at the LHC and ILC will be able to measure all these
Higgs boson masses, couplings and decay modes \cite{Barger:2006dh}.
Clearly, $\eta$ and $\psi$ (especially $\psi$) model yield lightest
of $H, A$ among all the models considered. In course of collider
searches, these two models will be differentiated from the others by
their relatively light heavy-Higgs sector.

\section{Conclusion}
In this work we have studied the lightest Higgs boson mass in
UU(1)$^{\prime}$ models against various model parameters and
particle masses. The model possesses a number of distinctive
features not found in the MSSM: the presence of the heaviest Higgs
boson $H^{\prime}$ (in addition to $H$ and $A$ present in the MSSM,
all studied in detail in  Fig. \ref{fig3}) as well as the
$\mu_{eff}$ dependencies of the sfermion masses (studied in Figs.
\ref{fig1} and \ref{fig2}). Concerning LEP Higgs measurements, it is
known that, bounds on the lightest Higgs boson in U(1)$^\prime$
extensions are similar to that of the MSSM, but its upper bound is
relaxed \cite{Barger:2006dh}. We have found rather generically that
the LEP bounds constrain all four models we have considered. The
Tevatron bounds, on the other hand, become effective for the $X$
model, primarily. These are felt also by the $\psi$ and $N$ models (
to a lesser extent than the $X$ model); however, the $\eta$ model
yields fundamentally light $h$ boson whose mass never nears the
Tevatron forbidden band. Nevertheless, one concludes from the
remaining three models that, the Tevatron bounds generically divide
all model parameters in two disjoint ranges: those pertaining to
low-mass domain and those to high-mass domain. For instance, the
Higgsino Yukawa coupling $h_s$, as seen from Fig. \ref{fig1},
requires large (close to unity) values to elevate $m_h$ above the
Tevatron's upper limit {\it i. e.} $\sim 168\ {\rm GeV}$. This kind
of restriction is seen also for other parameters, especially, the
U(1)$^{\prime}$ gauge coupling $g_Y^{\prime}$ (which needs to take
large values close to $2 g_Y$ to push $m_h$ in the Tevatron
territory in the models stemming from E(6) breaking).

In any case, at least for the parameter ranges considered, one achieves at the
firm conclusion that the Tevatron bounds can rule out certain portions of
the parameter space (as can be seen specially from Fig. \ref{fig2}). Of course,
this is in accord with the case whether $m_h$ is lying above or below the Tevatron
exclusion limits. For instance, if $m_h\sim\,168$ or higher then
Higgsino Yukawa coupling should be larger than $0.6$, for $m_h\sim\,159$ or
lower than this, Yukawa coupling of the singlet should be $0.5$ or
smaller according to our $X$ model. Besides this we observe that,
certain UU$^\prime$ models such as the $\eta$ model can be the first
one to be ruled out since its $m_h$ prediction is well below the
Tevatron exclusion limits, even with a unrealistically enhanced (close
to unity) gauge coupling $g^\prime_Y$.

Concerning the stop masses, we found that $X$
and E(6)-borrowed $N,\psi$ models are highly sensitive to Tevatron
(and any other collider bound) than in the MSSM due to the fact that
$\mu_{eff}$ determines not only the $LR$ (as in the MSSM) but also
the $LL$ and $RR$ (unlike the MSSM) entries of the stop and
sbottom mass-squared matrices. According to the model $X$, rule-outs of
stop searches can help to determine whether the lightest Higgs boson
is lying below or above the Tevatron Higgs mass measurements. Interestingly, low values of
$m_{\tilde{t}_1}$ can help to narrow down the range of $m_h$.
On the contrary, E(6)-based models can serve for the same aim,
but with the opposite behavior. This is another important signature of the model-dependence
surviving in UU(1)$^{\prime}$ models.

Another interesting aspect observed within the models considered is
that each model can predict a sensible splitting  among $m_A$ and
$m_H$ at varying order again in a model dependent fashion. In our
examples, their masses are generally larger than 500 GeV, and hence,
decoupled from the lightest Higgs (especially in $X$ and $N$ models).
Additionally, their mass splittings can be as large as tens of GeVs
in any model (much larger in $\eta$ and $\psi$ models). These observations
also hold for splittings between $Z^\prime$ and $H^\prime$ masses.

The results found above, though unavoidably carry a degree of model dependence,
can be directly tested at the LHC (and at the ILC with much higher precision). Measurements
of the Higgs mass at the LHC, if turn out to prefer large values like $130-140\ {\rm GeV}$ or
above, can be interpreted as preferring extensions of the MSSM like UU(1)$^{\prime}$
models. Depending on the future exclusion limits, one might find more regions of
parameter space excluded. For instance, if the Tevatron exclusion band widens down to $140\ {\rm GeV}$
border smaller and smaller values of $h_s$ and $g_{Y^{\prime}}$ become relevant. This
limit also forces the remaining heavy Higgs bosons to decouple from the light spectrum.
The plots presented in the figures are sufficiently ranged to cover possible
developments in future exclusion limits (which may come form continuing analysis of
the Tevatron data or from the early LHC data).

\textbf{Acknowledgements} The works of D. D. and H. S. were
partially supported by the Turkish Atomic Energy Agency (TAEK) via
CERN-CMS Research Grant, and by the IYTE-BAP project.


% \end{thebibliography}

\end{document}